\documentclass[11pt,letterpaper]{article}
\usepackage[utf8]{inputenc}
\DeclareUnicodeCharacter{03B8}{\circ}
\usepackage{graphicx}
\usepackage{scalerel}
\usepackage{amsmath}
\usepackage{booktabs}
\usepackage{tabu}
\usepackage{graphicx}
\usepackage{caption}
\usepackage{subcaption}
\usepackage{cite}
\usepackage{secdot}
\usepackage[margin=0.8in]{geometry}
\usepackage{amssymb}
\usepackage{url}
\usepackage[colorlinks = true,
            linkcolor = blue,
            urlcolor  = blue,
            citecolor = blue,
            anchorcolor = blue]{hyperref}\usepackage{ragged2e}
\usepackage{algorithm2e}
\usepackage[affil-it]{authblk}
\usepackage{abstract}

\usepackage[nodisplayskipstretch]{setspace}
\providecommand{\keywords}[1]
{
  \small	
  \textbf{\textit{Keywords---}} #1
}
\begin{document} 

\title{\Large{\textbf{
  Relic Abundance of Dark Matter with $\Delta$(54) Flavor Symmetry
 \vspace{-0.45em}}}}

\author{Hrishi Bora$^1$%
\thanks{\href{mailto:hrishi@tezu.ernet.in}{hrishi@tezu.ernet.in} (Corresponding author)}}
  
   \author{Ng. K. Francis$^1$\thanks{\href{mailto:francis@tezu.ernet.in}{francis@tezu.ernet.in}}}
  \author{Shawan Kumar Jha$^2$\thanks{\href{mailto:francis@tezu.ernet.in}{shawankumar@iitg.ac.in}}}

\affil{\vspace{-1.05em} $^1$Department of Physics, Tezpur University, Tezpur - 784028, India}
\affil{\vspace{-1.05em}$^2$Department of Physics, Indian Institute of Technology Guwahati,
Guwahati 781039, India}

\date{\vspace{-5ex}}
\maketitle
\begin{center}
\textbf{\large{Abstract}}\\
\justify

In this work we discuss the neutrino phenomenology in a $\Delta(54)$ discrete flavor symmetry  and evaluate
the relic abundance of dark matter (DM) and active neutrino-DM mixing angle  considering various cosmological
constraints.  We introduced two Standard Model Higgs particles along with vector-like fermions and   a new particle (S) which is a gauge singlet in the Standard Model. This modification leads to a mass matrix that diverges from the tribimaximal neutrino mixing pattern resulting in a non-zero reactor angle ($\theta_{13}$).
We also incorporated a $ Z_2 \otimes Z_3 \otimes Z_4$ symmetry for the specific interactions in our model.
The additional particle S
is a sterile neutrino which is considered to be a probable dark matter
candidate with mass in keV range.
\end{center}

\keywords{Dark matter; Inverse Seesaw; Tribimaximal neutrino mixing; Gauge singlet; Higgs particle }

\hspace{-1.9em} PACS numbers: 12.60.-i, 14.60.Pq, 14.60.St
\newpage
\section{Introduction}
\label{sec:intro}

The observation of neutrino masses and their flavor mixing, as evidenced by neutrino oscillations, raises a question regarding the origin of their  masses. \cite{aker2019improved, faessler2020status, araki2005measurement, cao2021physics, nath2021detection}. It seems extremely unlikely that neutrino masses function in the same way as the masses of charged fermions as the standard model does not contain right-handed neutrinos, in contrast to other fermions. Several models outside of the standard model (BSM) provide explanation for the the formation of neutrino masses, including the  seesaw mechanism \cite{minkowski1977mu, king2005testing, mohapatra1986mechanism}, radiative seesaw mechanism \cite{ma2009radiative}, extra-dimensional models \cite{mohapatra1999neutrino, arkani2001neutrino } and others. These references includes many current reviews on neutrino physics\cite{Nguyen:2018rlb, King:2014nza,King:2003jb,Cao:2020ans, Ahn:2014zja, mcdonald2016nobel, Nguyen:2020ehj, Hong:2022xjg, kajita2016nobel,de20212020,okada2021spontaneous, Ahn:2021ndu, PhongNguyen:2017meq,Buravov:2014dna, buravov2009elementary, barman2024neutrino, bora2024majorana, bora2023neutrino}.

This work introduces a method based on the $\Delta(54)$ flavour symmetry framework with the inverse seesaw mechanism.  Heterotic string models on factorizable orbifolds, such as the $T^2/Z_3$ orbifold, can exhibit the $\Delta$(54) symmetry. Within these string models, doublets are not detected as basic modes but only singlets and triplets are detected. However, in magnetized/intersecting D-brane models, doublets may develop into basic modes. The nice feature of $\Delta(54)$ flavor symmetry is that there are also doublets that can be used for the
quark sector. We can also suggest an extension to the Standard Model, utilizing $\Delta$(54) symmetry. To represent quarks in different ways, we may work with the singlets ($1_{1}, 1_{2}$) and doublets ($2_{1}, 2_{2}, 2_{3}, 2_{4} $) representations of $\Delta(54)$. This extension essentially includes the latest experimental data on a number of quark sector features, including three quark mixing angles, six quark masses, and the CP-violating phase \cite{vien2021extension}.

The neutrino phenomenology of the inverse see-saw is dependant upon the presence of an additional singlet fermion. According to this formalism, the lightest neutrino mass matrix is $M_{\nu} \approx M_d (M_T)^{-1} \mu M^{-1} {M_d}^T$, where M denotes the lepton number conserving interaction between right-handed and sterile fermions, and $M_d$ is the Dirac mass term.
Three more right-handed neutrinos and one gauge singlet chiral fermion field S as a sterile neutrino are added to the standard model particles in the Minimal Extended see-saw (MES), which is an extension of the canonical type-I see-saw.
The $4\times 4$ active-sterile neutrino mass matrix in this formalism is given by

\begin{equation}   \label{eq:1} 
M_{\nu}^{4\times4}= - \begin{pmatrix}
     M_D M^{-1}_R M^T_D &  M_D M^{-1}_R M^T_S \\
     M_S (M^{-1}_R)^T M^T_D &   M_S M^{-1}_R M^T_S \\
    \end{pmatrix} 
\end{equation}

where $M_D$, $M_R$ and $M_S$ are the Dirac, Majorana and Sterile neutrino mass matrices. The  Sterile neutrino mass is given as $\label{eq3}
   m_s \simeq -M_S M^{-1}_R M^T_S$.

Neutrinos with sub-eV can be produced from $M_D$ at the electroweak scale, $M_R$ at the TeV scale, and $M_S$ at the keV scale.
There are other important lines of evidence for dark matter (DM), such as Fritz Zwicky's 1933 observations of galaxy clusters, gravitational lensing (which Zwicky 1937 proposed could allow galaxy clusters to act as gravitational lenses), galaxy rotation curves in 1970, cosmic microwave background, and the most recent cosmology data provided by the Planck satellite.
It is known that dark matter (DM) makes up around 27\% of the universe, which is almost five times more than baryonic matter, based on the most current data from the Planck spacecraft. According to reports, the current dark matter abundance is given as\cite{taoso2008dark}
\begin{equation*}
    \Omega_{DM} h^2 = 0.1187 \pm 0.0017
\end{equation*}
The physics community has faced enormous challenges in its search for potential dark matter candidates with new mechanisms beyond the standard model. The important criteria that a particle must have in order to be taken into consideration as a viable DM candidate. All of the SM particles are not eligible to be DM candidates due to these restrictions. The particle physics community became motivated to investigate several BSM frameworks that may provide accurate DM phenomenology and can be evaluated in many experiments.

The paper is structured as follows: In sect \ref{sec: frame}, we introduce the $\Delta(54)$ discrete symmetry with inverse seesaw mechanism and discuss the characteristics of the flavor group relevant to constructing the model. In sect \ref{num}, we outline the allowable range for model parameters based on the constraints imposed by the $3{\sigma}$ range of neutrino oscillation data. In sect \ref{Res}, we summarize the sterile dark matter and Ly-$\alpha$ constraints. In sect \ref{conc}, we conclude our study and provide numerical results  within the model.

\section{Stucture of the Model}
\label{sec: frame}
For the implementation of the Inverse Seesaw mechanism, the fermion sector must be extended within the Standard Model framework.  We have introduced vector-like (VL) fermion denoted as $N$ which have the property of being gauge singlets inside the SM framework.  We also introduced a gauge singlet denoted as $S$, which is
considered to be a dark matter candidate in our model. In fact, the VEV of $\phi$ induces the mass term for the $S$ fermion.  We extended the model by a gauge singlet fermion. The main motivation
of this extended field is to incorporate dark matter phenomenology in our model. The $\Delta(54)$  group includes irreducible representations  $1_1$, $1_2$, $2_1$, $2_2$, $2_3$, $2_4$, $3_{1(1)}$, $3_{1(2)}$, $3_{2(1)}$ and $3_{2(2)}$.  
\vspace{1cm}

\begin{table}[ht]
    \centering
    \scalebox{0.9}{
 {\begin{tabular}{c c c c  c c  c c c c c c c }
    \hline
       \textrm{Field}  &  L & $l $ & $H$ & $H^{\prime}$  & $N$   & $S$ & $\chi$ & $\chi^\prime$ & $\zeta$  & $\xi$ & $\Phi_{S}$ & $\phi$  \\
     \hline
     \textrm{$\Delta(54)$}  &  $3_{1(1)}$ &  $3_{2(2)} $ & $1_{1}$ & $1_{2}$ & $3_{1(1)}$   & $3_{2(2)}$ & $1_2$ & $2_1$ & $1_{2}$  & $3_{2(1)}$ & $3_{1(1)}$ & $3_{1(2)}$ \\
     \textrm{Z}$_2$  &  1 & -1 & 1 & 1 & -1 & 1 & -1 & -1 & -1  & -1 & -1 & 1 \\
    \textrm{Z}$_3$  &  $\omega$ & $\omega$ & 1 & 1  & 1   & 1 & 1 & 1& 1  & $\omega$ & $\omega$ & 1 \\
\textrm{Z}$_4$  &  1 & -1 & 1 & 1  & 1   & 1 & -1 &-1& 1 & 1 & 1 & 1 \\
\textrm{U(1)}  &  1 & 1 & 0 & 0  & 1  & 1 & 0 & 0 & 0  & 0 & 0& 0 \\
     \hline
    \end{tabular}}}
    \caption{Particle content of our model}
    \label{tab:1}
    \end{table}

 We  developed a model based on the $\Delta(54)$ disrete symmetry with inclusion of additional flavons namely $\chi$, $\chi^\prime$,  $\zeta$, $\zeta^\prime$, $\xi$,  $\Phi_{S}$ and $\phi$. In order to avoid undesired interactions, we added additional symmetry $Z_2\otimes Z_3 \otimes Z_4$. Details on the particle  charge assignment and composition according to the flavor group are given in Table \ref{tab:1}. The left-handed leptons doublets and the right-handed charged lepton are assigned using the triplet representation of $\Delta(54)$.

The Lagrangian is as follows   :
 \begin{align}
  \mathcal{L} = & \frac{y_1}{\Lambda} ( l \Bar{L} ) \chi H + \frac{y_2}{\Lambda} ( l \Bar{L} ) \chi^\prime H   + \frac{\Bar{L} \Tilde{H^{\prime}} N}{\Lambda}y_{\xi} \xi + \frac{\Bar{L} \Tilde{H} N}{\Lambda} y_{s} \Phi_{s} + \frac{\Bar{L} \Tilde{H^{\prime}} N}{\Lambda} y_{a} \Phi_{s}  \nonumber  \\                 
 & + y_{\scaleto{NS}{4pt}}  \Bar{N^{c}}S  \zeta  + \frac{y_{s_1}}{\Lambda^2}\Bar{S} S^{c} \phi 
 \label{eq1}
\end{align}

The VEVs are  naturally considered as,
\begin{align*}
\langle \chi \rangle& =(v_{\chi})&
\langle \chi^\prime \rangle& =(v_{{\chi}^\prime},v_{{\chi}^\prime})&
\langle \Phi_S \rangle& =(v_s ,v_s,v_s)&
\langle \xi \rangle& =(v_{\xi}, v_{\xi}, v_{\xi})&\\
\langle \phi \rangle& = (v_{\phi},v_{\phi},v_{\phi})&
\langle \zeta \rangle& =(v_{\zeta})  
\end{align*}

The charged lepton mass matrix is given as \cite{ishimori2009lepton} 
\begin{align*}
    M_l= \frac{y_1 v}  {\Lambda}
    \begin{pmatrix}
    v_{\chi} & 0 & 0\\
    0 & v_{\chi}  & 0\\
    0 & 0 &  v_{\chi} 
    \end{pmatrix}  +  
    \frac{y_2 v}  {\Lambda}
    \begin{pmatrix}
    -\omega v_{\chi^\prime} + v_{\chi^\prime} & 0 & 0\\
    0 & -\omega^2 v_{\chi^\prime} + \omega^2 v_{\chi^\prime}  & 0\\
    0 & 0 &   -v_{\chi^\prime} + \omega v_{\chi^\prime}
    \end{pmatrix} 
\end{align*}

where $y_1$ and $y_2$ are coupling constants.

\subsection{Effective neutrino mass matrix}

The Lagrangian may be used to construct the mass matrices related with the neutrino sector after applying $\Delta(54)$ and electroweak symmetry breaking. The $M_{S}$ scale is an essential assumption of the ISS theory, providing small neutrino masses. To reduce the right-handed neutrino masses to the TeV scale, the $M_{S}$ scale must be at the keV level. The inverse seesaw model is a TeV-scale seesaw model that permits heavy neutrinos remain as light as a TeV and Dirac masses to be as large as those of charged leptons, while maintaining compatibility with light neutrino masses in the sub-eV range.

\begin{equation}
M_{NS}=   y_{\scaleto{NS}{3pt}} \begin{pmatrix}
     v_{\zeta} & 0 & 0\\
    0 &  v_{\zeta} & 0\\
    0 & 0 &  v_{\zeta}
    \end{pmatrix}
\end{equation}

 \begin{equation}
 M_{\nu N}= \frac{v}{\Lambda}        \begin{pmatrix} 
    y_{\scaleto{\xi}{6pt}} v_{\scaleto{\xi}{6pt}} & y_{s}v_{s} + y_{a} v_a & y_{s}v_{s} - y_{a} v_a\\
   y_{s}v_{s} - y_{a} v_a &   y_{\scaleto{\xi}{6pt}} v_{\scaleto{\xi}{6pt}} & y_{s}v_{s} + y_{a} v_a\\
    y_{s}v_{s} + y_{a} v_a & y_{s}v_{s} - y_{a} v_a &   y_{\scaleto{\xi}{6pt}} v_{\scaleto{\xi}{6pt}}
    \end{pmatrix}
   \end{equation}

The Sterile neutrino mass matrix can be written as,
\begin{equation}    
M_{S}=  \frac{y_{s_1}}{\Lambda^{2}}\begin{pmatrix}
     v_{\phi} & 0 & 0\\
    0 &  v_{\phi} & 0\\
    0 & 0 &   v_{\phi}
    \end{pmatrix} 
\end{equation}

In the Inverse seesaw framework, the effective neutrino mass matrix can be written as
\begin{equation} 
m_\nu = M_{\nu N}M_{NS}^{-1}M_{S}M_{NS}^{-1}M^\prime_{\nu N}
\label{eq6}
\end{equation}

The resultant mass matrix from Eq. (\ref{eq6})
\begin{equation}
\label{eq10}
    m_\nu= v^2
    \begin{pmatrix}
     \frac{s}{M^2}(2a^{2} + 2c^{2} + x^{2})   &   \frac{s}{M^2}(-a^{2} + c^{2} + 2c x^{})   &    \frac{s}{M^2} (-a^{2} + c^{2} + 2c x)\\
     \frac{s}{M^2}(-a^{2} + c^{2} + 2c x)    &  \frac{s}{M^2}(2a^{2} + 2c^{2} + x^{2})   &       \frac{s}{M^2}(-a^{2} + c^{2} + 2c x) \\
   \frac{s}{M^2}(-a^{2} + c^{2} + 2c x) &   \frac{s}{M^2}(-a^{2} + c^{2} + 2c x)  &    \frac{s}{M^2}(2a^{2} + 2c^{2} + x^{2})
    \end{pmatrix}
\end{equation}

where    $a = \frac{y_a v_sv}{\Lambda}$,  $c = \frac{y_s v_sv}{\Lambda}$,  $x = \frac{y_{\scaleto{\xi}{3pt}} v_{\scaleto{\xi}{3pt}}v}{\Lambda}$ , $s= \frac{y_{s_1} v_{\phi}}{\Lambda^2} $ and $M= y_{\scaleto{NS}{3pt}} v_\zeta  $. The dimension of the problem is absorbed by the term $v$ and the components of the matrix are unaffected . Phase redefinitions of the charged lepton fields can absorb the phase of $v$, allowing it to be considered as a real parameter without losing generality \cite{lei2020minimally}.\\

The $4 \times 4$ active sterile neutrino mass matrix represented as in Eq.\ref{eq:1} becomes,

\begin{equation*}
M_{\nu}^{4\times4}=    \begin{pmatrix}
\begin{pmatrix}
 -\frac{v^2(2a^{2} + 2c^{2} + x^{2})}{M\Lambda^2}   &   \frac{v^2(a^{2} - c^{2} + 2c x)}{M\Lambda^2}   &    \frac{v^2(a^{2} - c^{2} + 2c x)}{M\Lambda^2} \\
     \frac{v^2(a^{2} - c^{2} + 2cx)}{M\Lambda^2}    &   -\frac{v^2(2a^{2} + 2c^{2} + x^{2})}{M\Lambda^2}    &       \frac{a^{2} - c^{2} + 2c x}{M\Lambda^2} \\
   \frac{a^{2} - c^{2} + 2c x}{M\Lambda^2} &   \frac{a^{2} - c^{2} + 2c x}{M\Lambda^2}  &    -\frac{v^2(2a^{2} + 2c^{2} + x^{2})}{M\Lambda^2} 
   \end{pmatrix} &

\begin{pmatrix}
    -\frac{svx}{M} & -\frac{(a+c)sv}{M\Lambda} & \frac{(a-c)sv}{M\Lambda} \\
    \frac{(a-c)sv}{M\Lambda} & -\frac{svx}{M\Lambda} & -\frac{(a+c)sv}{M\Lambda}\\
    -\frac{(a+c)sv}{M\Lambda} & \frac{(a-c)sv}{M\Lambda} &  -\frac{svx}{M\Lambda}   
    \end{pmatrix} \\

\begin{pmatrix}
    -\frac{svx}{M \Lambda} & \frac{(a-c)sv}{M\Lambda} & -\frac{(a+c)s}{M\Lambda} \\
    -\frac{(a+c)s}{M\Lambda} & -\frac{svx}{M\Lambda} & -\frac{(a-c)sv}{M\Lambda}\\
    \frac{(a-c)sv}{M\Lambda} & -\frac{(a+c)sv}{M\Lambda} &  -\frac{svx}{M\Lambda}   
    \end{pmatrix}  &
\begin{pmatrix}
   - \frac{s^2}{M} & 0 &\\
    0 & -\frac{s^2}{M}&  0\\
    0 & 0 &  -\frac{s^2}{M}
\end{pmatrix}
  \end{pmatrix}
\end{equation*}

 The final neutrino mixing matrix for the active-sterile mixing takes $4 \times 4$
form as,

\begin{equation} \label{eq8}
V \simeq  \begin{pmatrix} (1-\frac{1}{2} R R^\dagger)U & R\\
    -R^\dagger U &  1-\frac{1}{2} R R^\dagger
\end{pmatrix}   
\end{equation}

where $R =   M_D M^{-1}_R M^T_S ( M_S M^{-1}_R M^T_S)$ is a $3\times 3$ matrix representing the strength of active
sterile mixing and $U$ is the leptonic mass matrix for active neutrinos \cite{zhang2012light}. 
\begin{equation} R \simeq
    \begin{pmatrix}
        \frac{s^3 v x}{M^2 \Lambda} & \frac{s^3 v (a+c)}{M^2 \Lambda} & \frac{s^3 v (-a+c)}{M^2 \Lambda}\\
        \frac{s^3 v (-a+c)}{M^2 \Lambda}&\frac{s^3 v x}{M^2 \Lambda} & \frac{s^3 v (a+c)}{M^2 \Lambda}\\
        \frac{s^3 v (a+c)}{M^2 \Lambda}&\frac{s^3 v (-a+c)}{M^2 \Lambda}&\frac{s^3 v x}{M^2 \Lambda}
    \end{pmatrix}
\end{equation}
\noindent The neutrino mass matrix $m_\nu$ can be diagonalized by the PMNS matrix $U$ as
\begin{equation}
    U^\dagger m^{(i)}_\nu U^* = \textrm{diag(}m_1, m_2, m_3 \textrm{)}
\end{equation}
 We can analytically calculate $U$ using the relation $U^\dagger h U = \textrm{diag(}m_1^2, m_2^2, m_3^2 \textrm{)}$, where $h = m_\nu m^{\dagger}_\nu$. We followed the framework of Adhikary et al. \cite{adhikary2013masses} for  calculating oscillation parameters of a generalized neutrino mass
matrix. The row-wise elements of $U$ are given in terms of the elements
of the $h$ and its eigenvalues $m^2_i$.

\begin{equation}
       U_{1i} = \frac{(h_{22} - m^2_i)h_{13} - h_{12}h_{23}}{N_i}; \quad  U_{2i} = \frac{(h_{11} - m^2_i)h_{23} -h^{*}_{12}h_{13}}{N_i}; \quad
   U_{3i} = \frac{\lvert h_{12} \rvert^2 - (h_{11}-m^2_i)(h_{22}- m^2_i) }{N_i} 
\end{equation}

\section{Numerical Analysis} 
\label{num}

\noindent The
mass matrix in Eq. (\ref{eq10}) gives the effective neutrino mass
matrix in terms of the model complex parameters $a$, $c$, and
$x$. We can find the parameter values of the model by fitting the
model to the current neutrino oscillation data.  We
use the 3$\sigma$ interval for the neutrino oscillation parameters
( $\theta_{12}, \theta_{23}, \theta_{13}, \Delta m^2_{21}
, \Delta m^2_{31}$ ) as presented in Table \ref{tab:2}. A
 constraint was applied   on
the sum of absolute neutrino masses	from the cosmological bound $\sum_{i} m_i < 0.12 eV$. In our study, the three complex parameters of the model
are treated as free parameters and are allowed to run over the
following ranges:
$\lvert a \rvert \in [0, 1] eV $, $ \phi_a \in [-\pi , \pi]$ ;
$\lvert c \rvert \in [0, 10^{-1}] eV$ ; $\phi_c \in [-\pi , \pi]$ ;
$\lvert x \rvert \in [0, 10^{-3}] eV$ , $\phi_x \in [-\pi , \pi]$. The two real parameters are allowed to run over the ranges: $ M \in [10^{13}, 10^{14}] eV$, $ s \in [10^{8}, 10^{9}] eV$  \\
where $\phi_a$ , $\phi_c$ and $\phi_x$ are the phases.

\begin{table}[ht]
\centering
 \scalebox{1}{
  \begin{tabular}{ | l | c | r |}
    \hline
    Parameters & NH (3$\sigma$) & IH (3$\sigma$) \\ \hline
    $\Delta{m}^{2}_{21}[{10}^{-5}eV^{2}]$ & $6.82 \rightarrow 8.03$ & $6.82 \rightarrow 8.03$ \\ \hline
    $\Delta{m}^{2}_{31}[{10}^{-3}eV^{2}]$ & $2.428 \rightarrow 2.597$ & $-2.581 \rightarrow -2.408 $\\ \hline
    $\sin^{2}\theta_{12}$ & $0.270 \rightarrow 0.341$ & $0.270 \rightarrow 0.341$ \\ \hline
     $\sin^{2}\theta_{13}$ & $0.02029 \rightarrow 0.02391$ & $0.02047 \rightarrow 0.02396$ \\ \hline
    $\sin^{2}\theta_{23}$ & $0.406 \rightarrow 0.620$ & $0.410 \rightarrow 0.623$ \\ \hline
    $\delta^{\circ}_{CP}$ & $108 \rightarrow 404$ & $192 \rightarrow 360$ \\ \hline
  \end{tabular}}
  \caption{ The neutrino oscillation parameters from NuFIT 5.2 (2022) \cite{esteban2022nufit}}
    \label{tab:2}
\end{table}

\begin{figure}[ht]
     \centering
     \begin{subfigure}{0.45\textwidth}
         \centering
         \includegraphics[width=\textwidth]{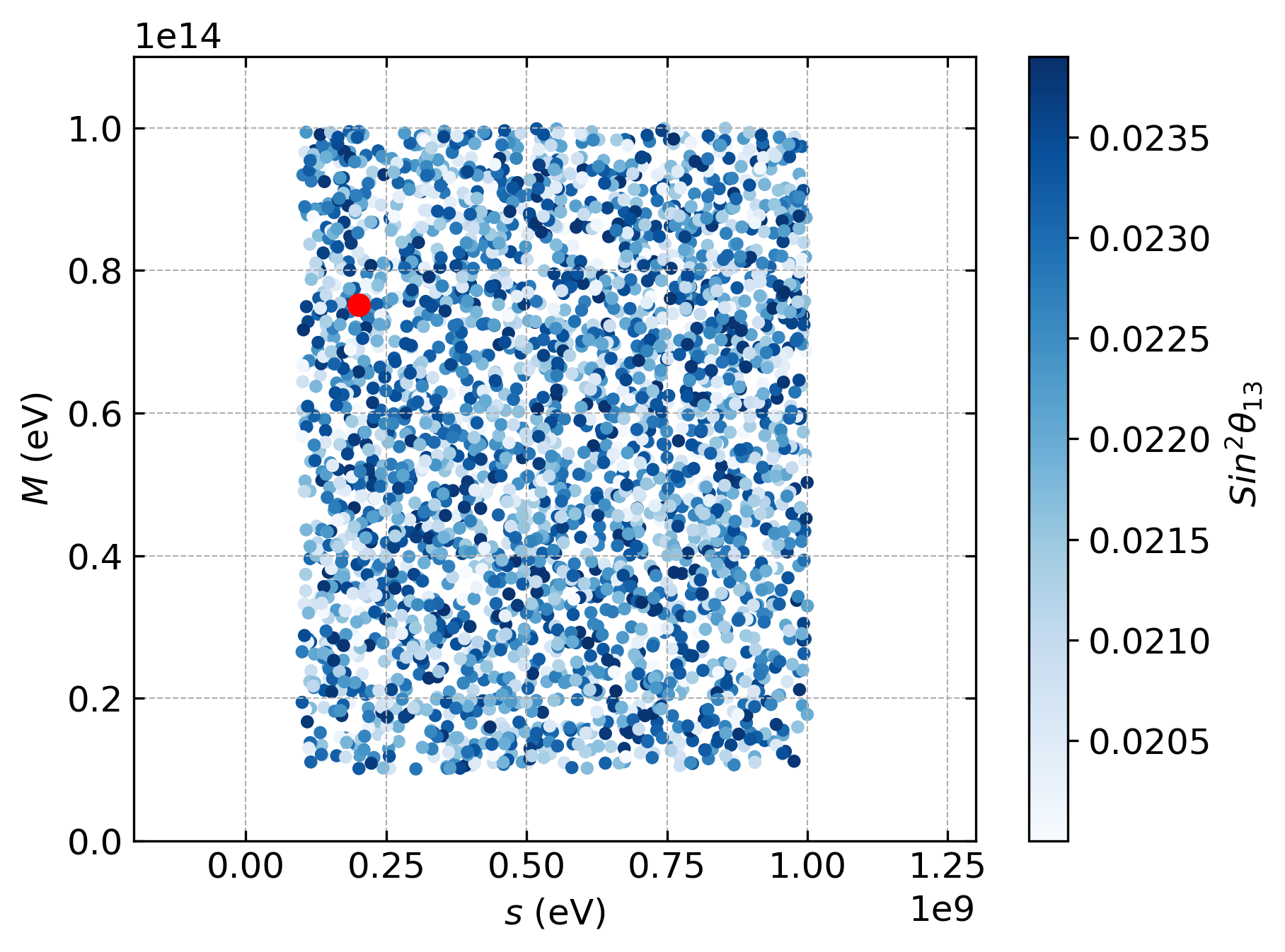}
     \end{subfigure}
     \hfill
     \begin{subfigure}{0.45\textwidth}
         \centering
         \includegraphics[width=\textwidth]{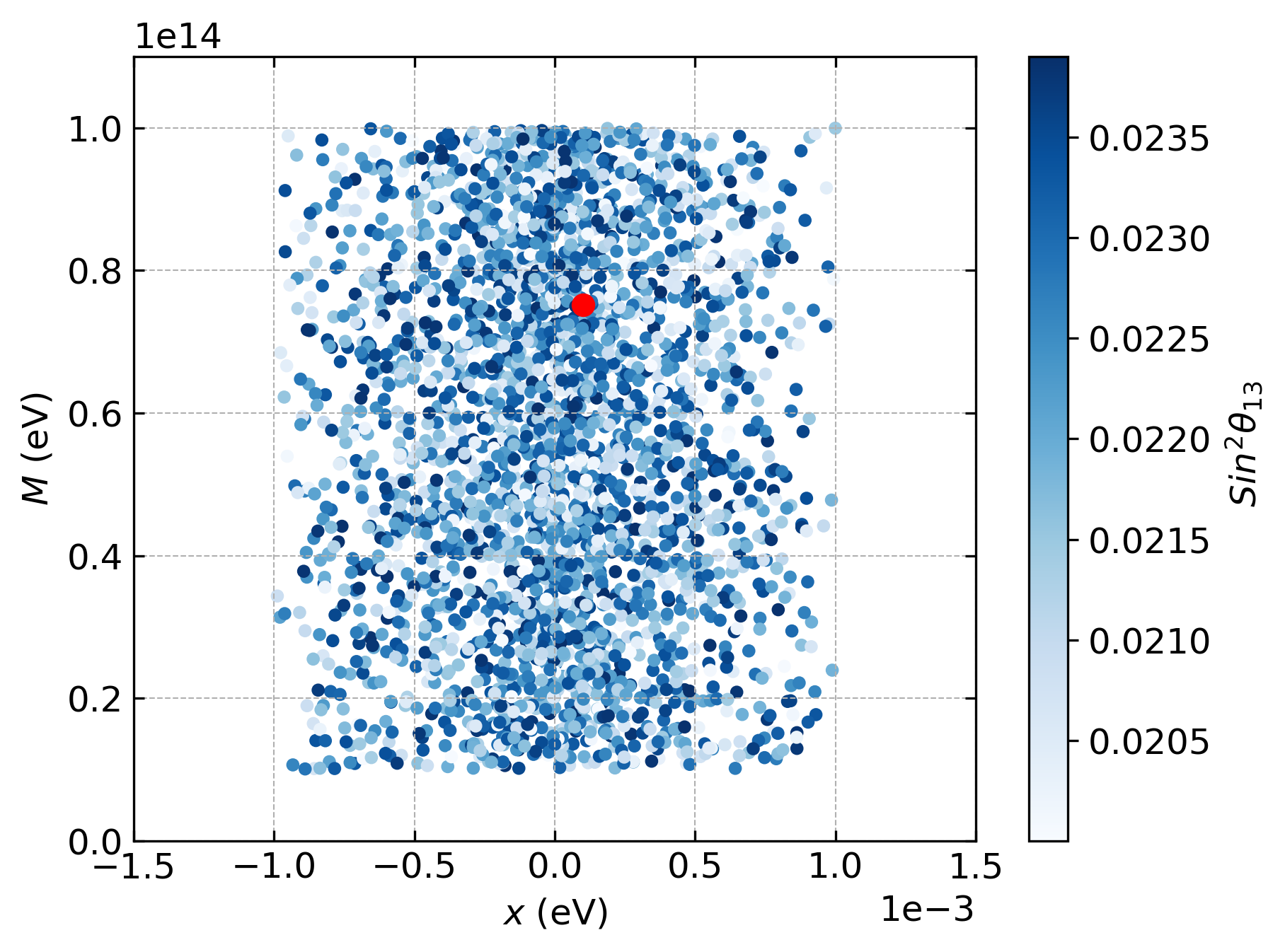}
     \end{subfigure}
     \centering
     \begin{subfigure}{0.45\textwidth}
         \centering

         \vspace{0.8cm}
         \includegraphics[width=\textwidth]{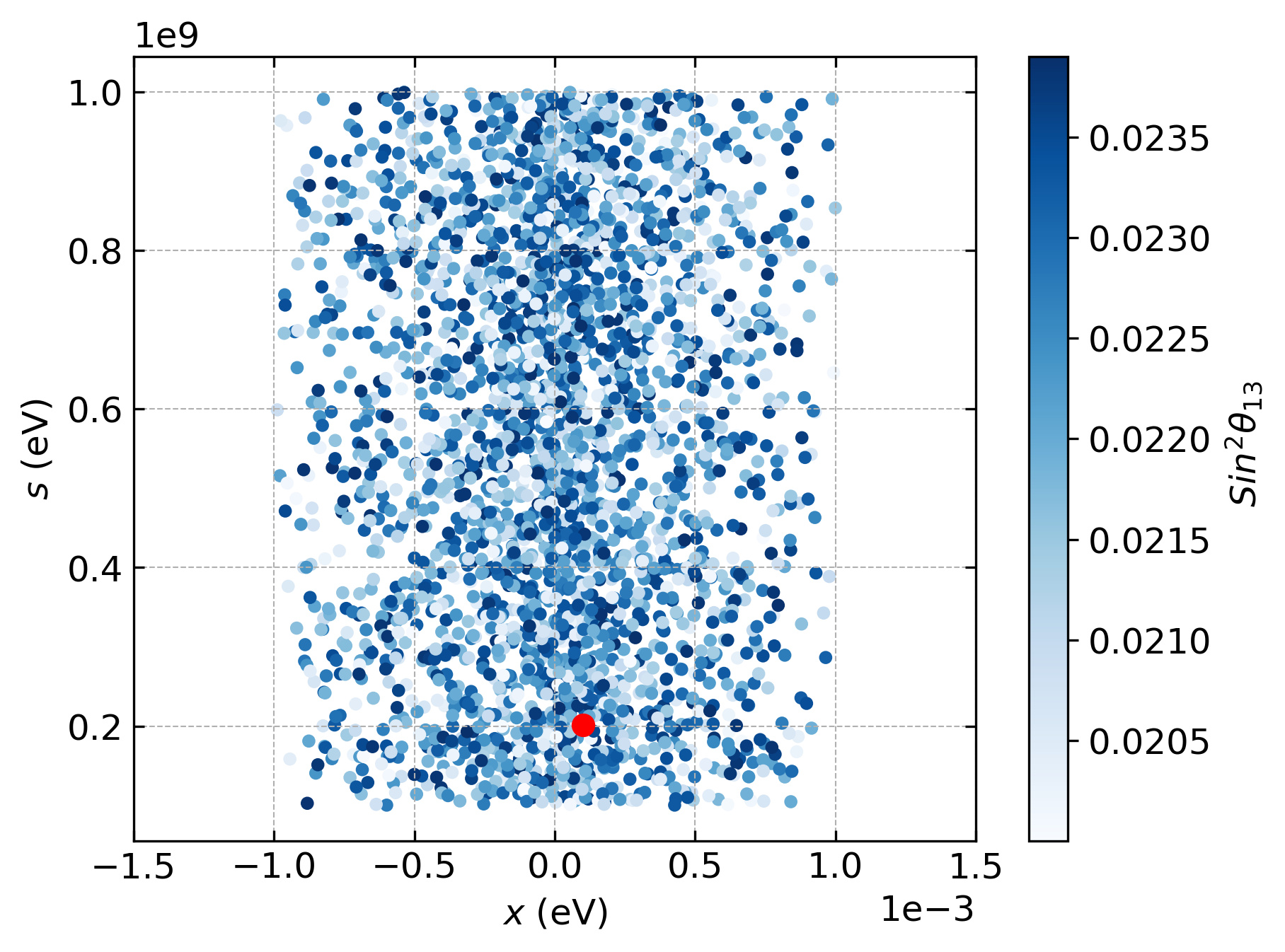}
     \end{subfigure}
     \hfill
     \begin{subfigure}{0.45\textwidth}
         \centering
         \includegraphics[width=\textwidth]{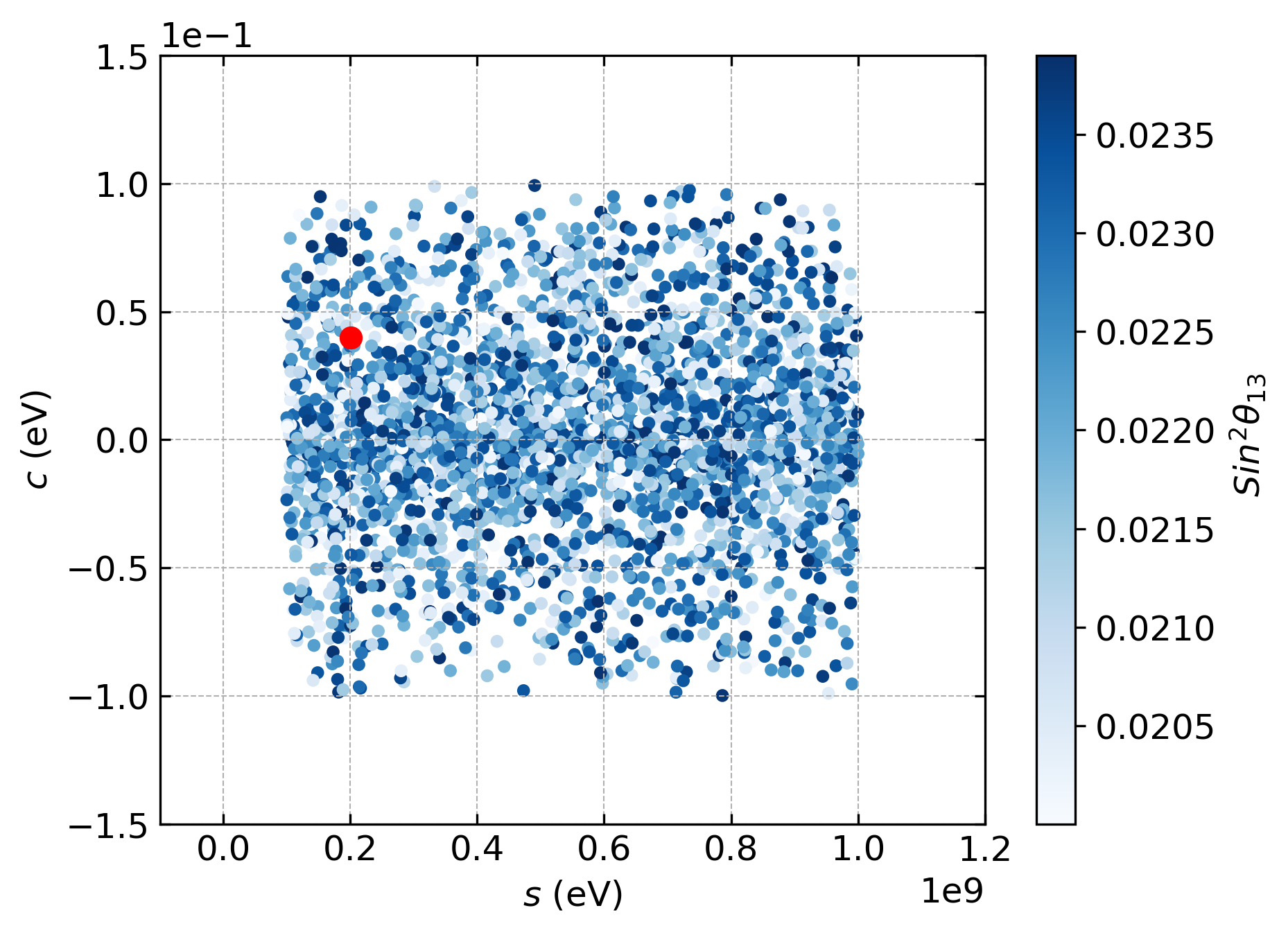}
     \end{subfigure}
      \caption{Correlation between the  parameters $x$, $s$, $a$, $M$ and $s$ . The red dot marker indicates the best-fit values corresponding to $\chi^2$ min.}
\end{figure}

The neutrino mass matrix $m_\nu$ can be diagonalized by the PMNS matrix $U$ as follow \cite{lei2020minimally}:
\begin{equation}
    \label{eq:12}
    U^\dagger m_\nu U^* = \textrm{diag(}m_1, m_2, m_3 \textrm{)}
\end{equation}

 We numerically calculated $U$ using the relation $U^\dagger h U = \textrm{diag(}m_1^2, m_2^2, m_3^2 \textrm{)}$, where $h = m_\nu m^{\dagger}_\nu$. The neutrino oscillation parameters $\theta_{12}$, $\theta_{13}$, $\theta_{23}$ and $\delta$ can be obtained from $U$ as 
\begin{equation}
    \label{eq:13}
    s_{12}^2 = \frac{\lvert U_{12}\rvert ^2}{1 - \lvert U_{13}\rvert ^2}, ~~~~~~ s_{13}^2 = \lvert U_{13}\rvert ^2, ~~~~~~ s_{23}^2 = \frac{\lvert U_{23}\rvert ^2}{1 - \lvert U_{13}\rvert ^2}
\end{equation}

and $\delta$ may be given by
\begin{equation}
    \label{eq:14}
    \delta = \textrm{sin}^{-1}\left(\frac{8 \, \textrm{Im(}h_{12}h_{23}h_{31}\textrm{)}}{P}\right)
\end{equation}
with 
\begin{equation}
    \label{eq:15}
     P = (m_3^2-m_2^2)(m_2^2-m_1^2)(m_3^2-m_1^2)\sin 2\theta_{12} \sin 2\theta_{23} \sin 2\theta_{13} \cos \theta_{13}
\end{equation}

We measured the difference between the neutrino mixing parameters and the latest experimental data by minimizing the resulting $\chi^2$ function, which allowed us to modify the $\Delta(54)$ model to fit the experimental data:

\begin{equation}
	\label{eq:16}
	\chi^2 = \sum_{i}\left(\frac{\lambda_i^{model} - \lambda_i^{expt}}{\Delta \lambda_i}\right)^2,
\end{equation}

where $\lambda_i^{model}$ is the $i^{th}$ observable predicted by the model, $\lambda_i^{expt}$ stands for  $i^{th}$ experimental best-fit value and $\Delta \lambda_i$ is the 1$\sigma$ range of the observable.

The best-fit values for $ x $, $ c $ and $ a $ obtained are $0.109 \times 10^{-3}$ eV, $0.401 \times 10^{-1}$ eV and $0.105 \times 10^{-1}$ eV respectively. The values of $M$ and $s$ are $0.752 \times 10^{14}$ eV and $0.201 \times 10^{9}$ eV respectively. These values are used further to calulate the neutrino oscillation parameters of our model.

Correspondingly, the best-fit values for
the neutrino oscillation parameters are, $\sin^2\theta_{12} = 0.31940$, $\sin^2 \theta_{13} = 0.02394$, $\sin^2 \theta_{23} =  0.51231 $, $\sin \delta_{CP} = 0.094$. The best-fit values for other parameters, such as  $\Delta m^2_{21}/\Delta m^2_{31}$ is 0.029, which corresponds to the $\chi^2$-minimum.

The Sterile mass is obtained as \cite{zhang2012light}:
\begin{equation}
    m_s = \frac{s^2}{M}
\end{equation}

which is considered as the dark matter mass and is represented as $M_{DM}$ in our analysis.
\vspace{0.8cm}

\begin{figure}[ht]
     \centering
     \begin{subfigure}{0.46\textwidth}
         \centering
         \includegraphics[width=\textwidth]{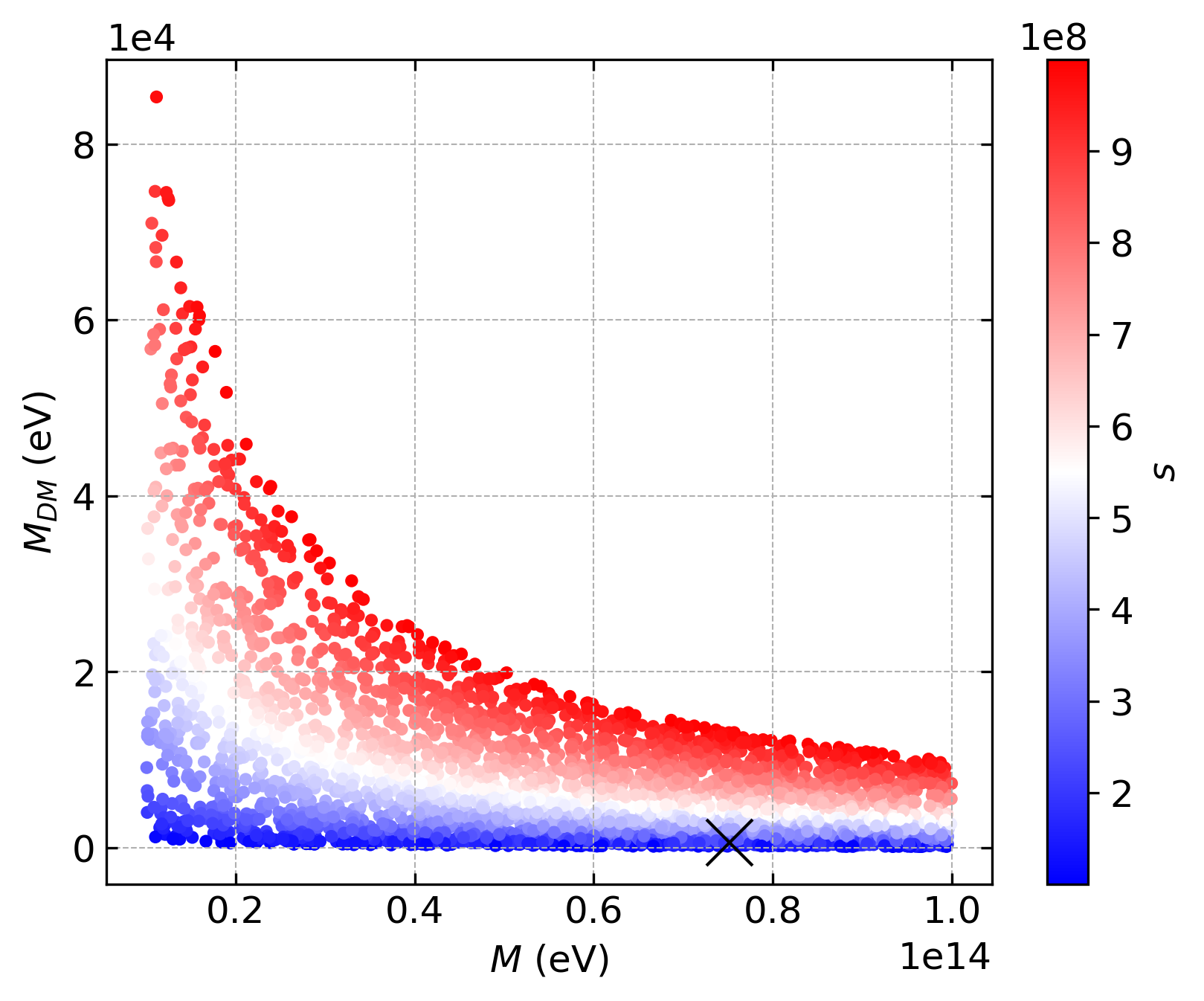}
     \end{subfigure}
     \hfill
     \begin{subfigure}{0.46\textwidth}
         \centering
         \includegraphics[width=\textwidth]{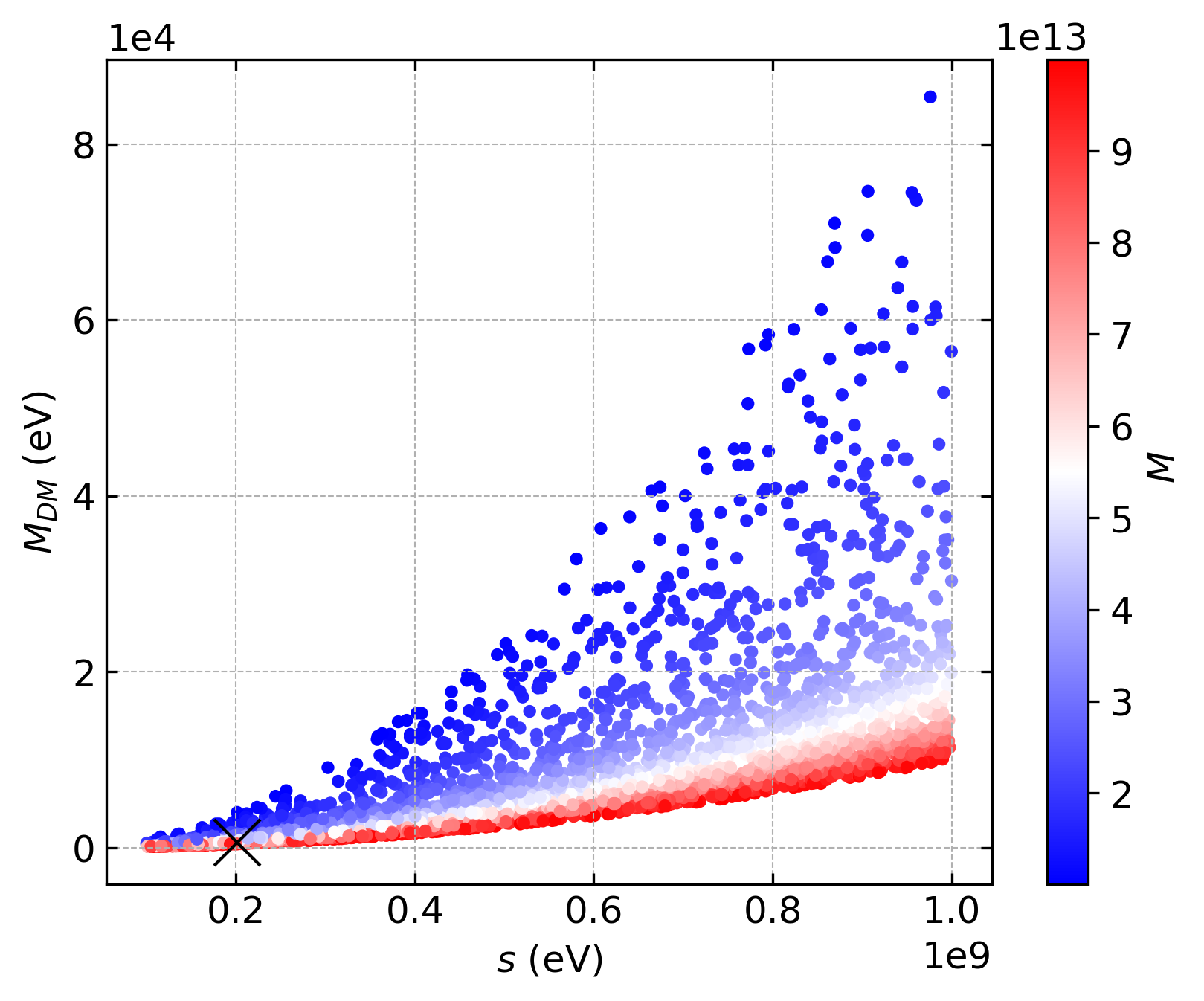}
     \end{subfigure}
    \caption{Allowed regions of the parameter $M_{DM}$ . The best fit values are indicated by cross marker.}
    \label{Fig2}
\end{figure}

\section{Sterile Dark Matter} 
\label{Res}

A non-resonantly created sterile dark matter (DM) may be easily accommodated within an extended SM of particle physics. Since the mass of the sterile neutrino is limited to a few keV, it is a strong candidate for warm dark matter. The only way the sterile neutrinos interact with the particles of the standard model is via mixing with the active neutrinos. Thus, the DM abundance can ultimately build up from the active neutrinos that are a part of the original plasma (because they interact weakly).  This mechanism is known as
the Dodelson and Widrow (DW) mechanism. 

When lepton asymmetry is absent, sterile neutrino non-resonant production (NRP) occurs. On the other hand, resonant sterile neutrinos with tiny mixing angles can be produced for a convincing amount of lepton asymmetry in the original plasma, leading to noticeably cooler momenta. Another name for this process is the resonant production mechanism, or Shi and Fuller (SF). Non-resonant production accounts for the smallest amount of dark matter contribution that can be created as a result of the dark matter mass and the mixing angle.
As a non-resonant DM candidate, we have taken into account a singlet fermion (S) in our model. Therefore $m_{DM}$ will be used to represent the fermion mass.

We solve the  parameters of the model with some fixed values of variables such as the $M$
in the range $10^{13}- 10^{14}$ GeV. Additionally, the gauge singlet $S$ is considered with a mass between $10^{8}$ and $10^{9}$ eV. By calculating these values, we are able to obtain the mixing angle $sin^2(2\theta_{DM})$ that satisfies the cosmological constraints as well as the required mass $m_{DM}$ in the keV range. In our study, the non-vanishing components  $V_{14}$ and $V_{34}$ of the mixing matrix V, contribute to the mixing angles. For any species, the relic abundance may be expressed as \cite{kolb1990early},

\begin{equation}
    \Omega h^2 = \frac{\rho_{x_o}}{\rho_{crit}} = \frac{s_o Y_{\infty} m}{\rho_{crit}}
\end{equation}

where, $\rho_{crit}$ is the critical energy density of the
universe, $\rho_{x_o}$ is present energy density of $x$, $Y_{\infty}$ is the present abundance of the particle $x$ and $s_{o}$ is the present day entropy.
Also we can get the values of $\rho_{crit} \approx 1.054 \times 10^{-5}$ h$^2$ GeV cm$^{-3}$ and $s_{o} \approx 2886$ cm$^{-3}$ from
Particle Data Group (PDG). For sterile neutrinos, it can be expressed as\cite{asaka2007lightest}:

\begin{equation}
    \Omega_{\alpha x} = \frac{m_x Y_{\alpha x}}{3.65 \times 10^{-9} \text{h$^2$} \text{GeV}}
\end{equation}

where $\alpha = e, \mu, \tau$ .
According to the active-sterile mixing and the sterile mass, which is proportional to the resultant relic abundance of a sterile neutrino state with a non-vanishing mixing to the active neutrinos is expressed as
 \cite{abada2014dark, ng2019new},
\begin{equation}
    \Omega_{\alpha S} h^2 = 1.1 \times 10^7
\sum
C_{\alpha}(m_s)\lvert V_{\alpha s}\rvert^2(
\frac{m_s}{\text{keV}} )^2
\end{equation}
where,
\begin{equation}
    C_{\alpha} = 2.49 \times 10^{-5} \frac{Y_{\alpha s} \text{keV}}{sin^2(\theta_{\alpha s})m_s}
\end{equation}

Using the solution of the Boltzmann equation \cite{gautam2020phenomenology}, $C_\alpha$, the active flavor-dependent coefficients, may be determined numerically.
We replace $s$ with $DM$  in the following formula for relic abundance, taking into account the sterile neutrino as a potential dark matter  candidate and using the parametrization $\lvert V_{\alpha s} \simeq sin(\theta_{\alpha s})\rvert$. Consequently, the relic abundance simplified equation for non-resonantly generated dark matter takes the form
 \cite{asaka2007lightest, abazajian2001sterile}:

\begin{equation}
   \Omega_{DM} h^2 \simeq 0.3 \times 10^{10} Sin^2(2\theta_{DM})(\frac{M_{DM} \times 10^{-2}}{\text{keV}})^2
\end{equation}

where, $\Omega_{DM}$ is directly proportional to $m_{DM}$ which is the DM mass as mentioned earlier
and $sin^2(2\theta_{DM})$ viz the active-DM mixing angle with $sin^2(2\theta_{DM}) = 4(V^2_{14} + V^2_{34})$. Here, $V^2_{14}$ and $V^2_{34}$ are the $14$ and $34$ element of the $V$ matrix in Eq.\ref{eq8}.
\vspace{1cm}

\begin{figure}[ht]
     \centering
     \begin{subfigure}{0.46\textwidth}
         \centering
         \includegraphics[width=0.9\textwidth]{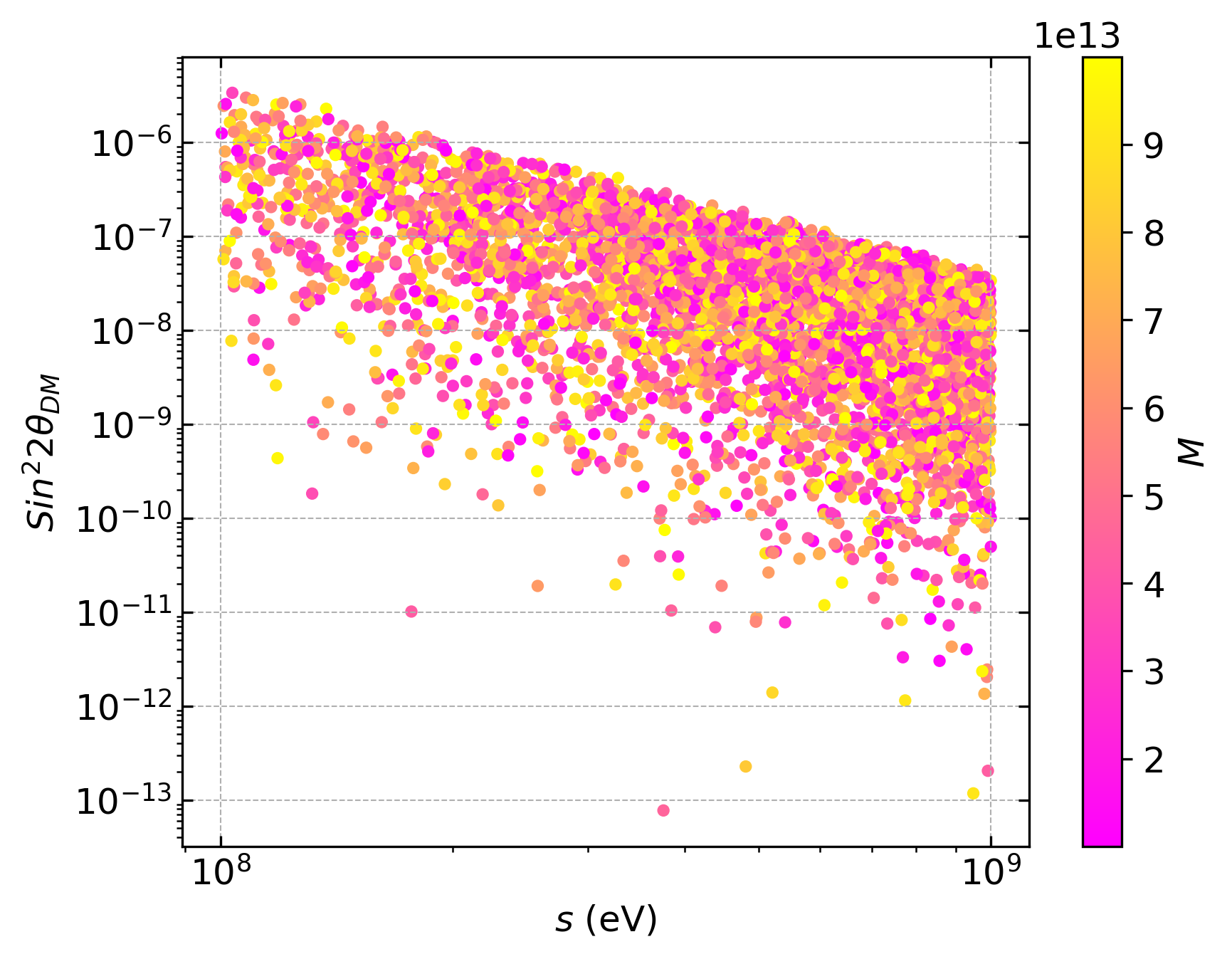}
     \end{subfigure}
     \hfill
     \begin{subfigure}{0.46\textwidth}
         \centering
         \includegraphics[width=0.9\textwidth]{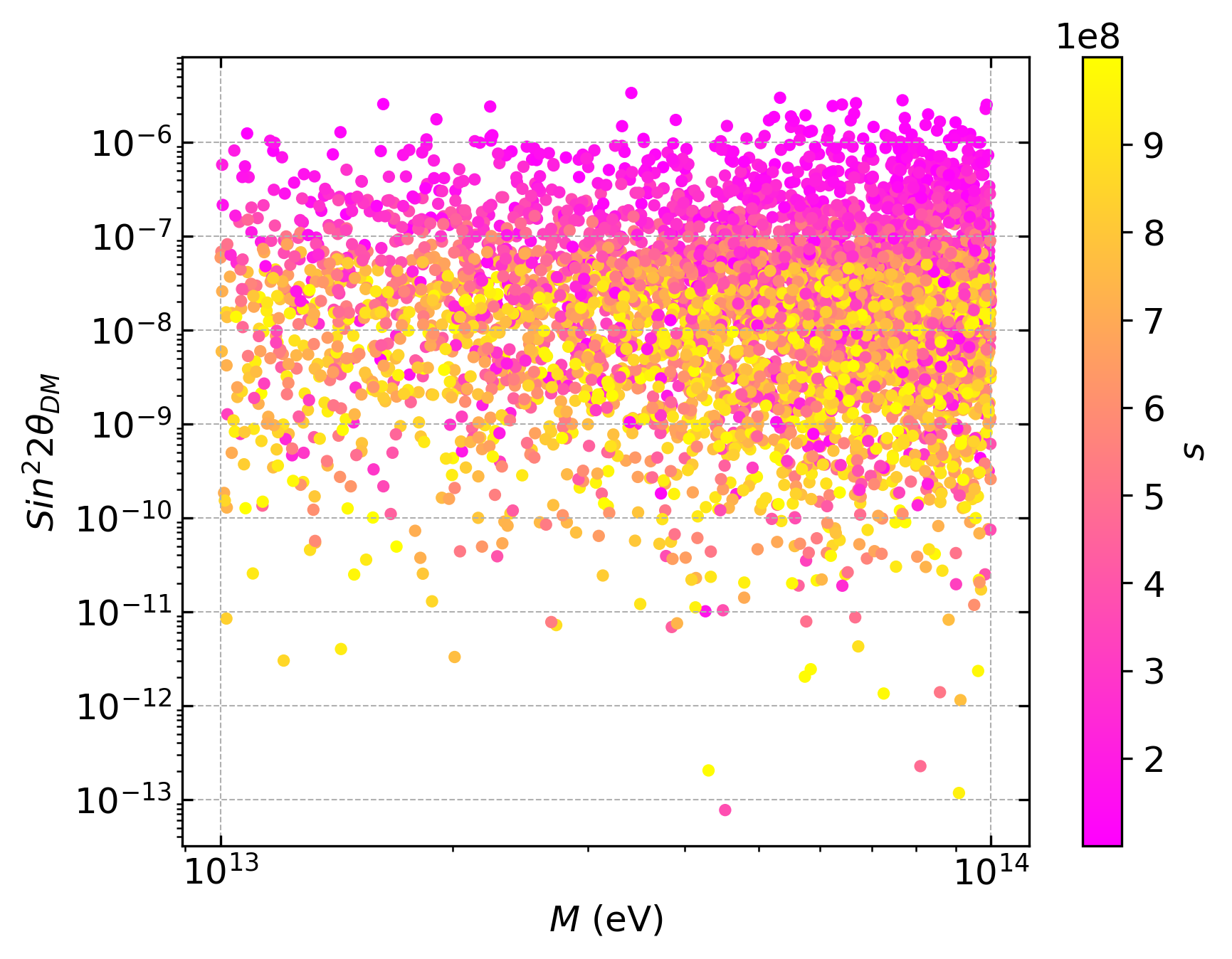}
     \end{subfigure}
     \end{figure}
\begin{figure}[h]
\centering
\includegraphics[width=0.4\textwidth]{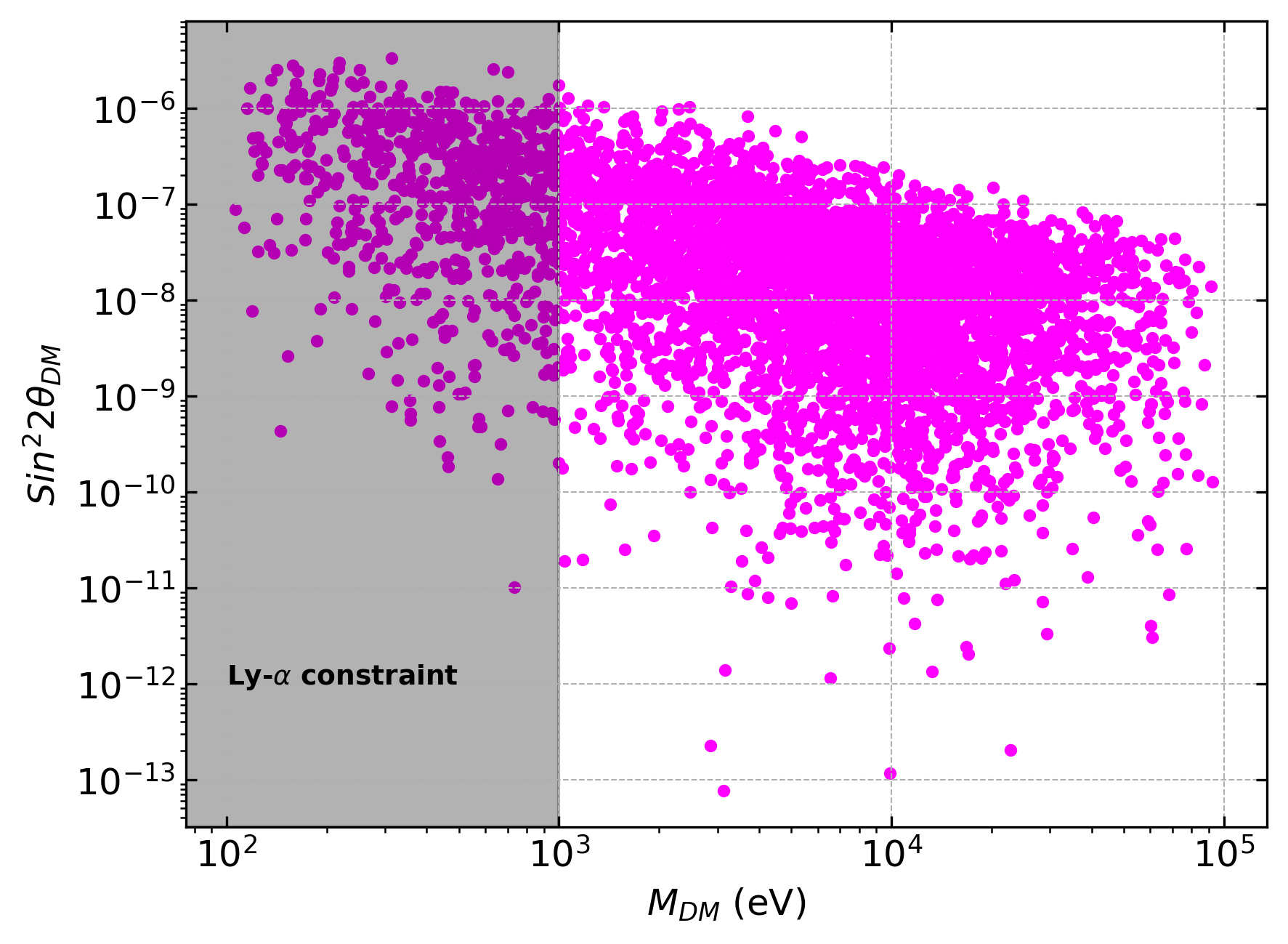}
    \caption{Plot in the first row shows active neutrino-DM mixing angle ($sin^2\theta_{DM}$) as a function of Gauge singlet mass ($s$) and RHN mass ($M$) respectively, in the second row shows as a function of dark matter mass($M_{DM}$)  and  including constraints
from Lyman-$\alpha$ for NH.}
    \label{Fig5}   
\end{figure}

\begin{figure}[ht]
     \centering
     \begin{subfigure}{0.46\textwidth}
         \centering
\includegraphics[width=0.9\textwidth]{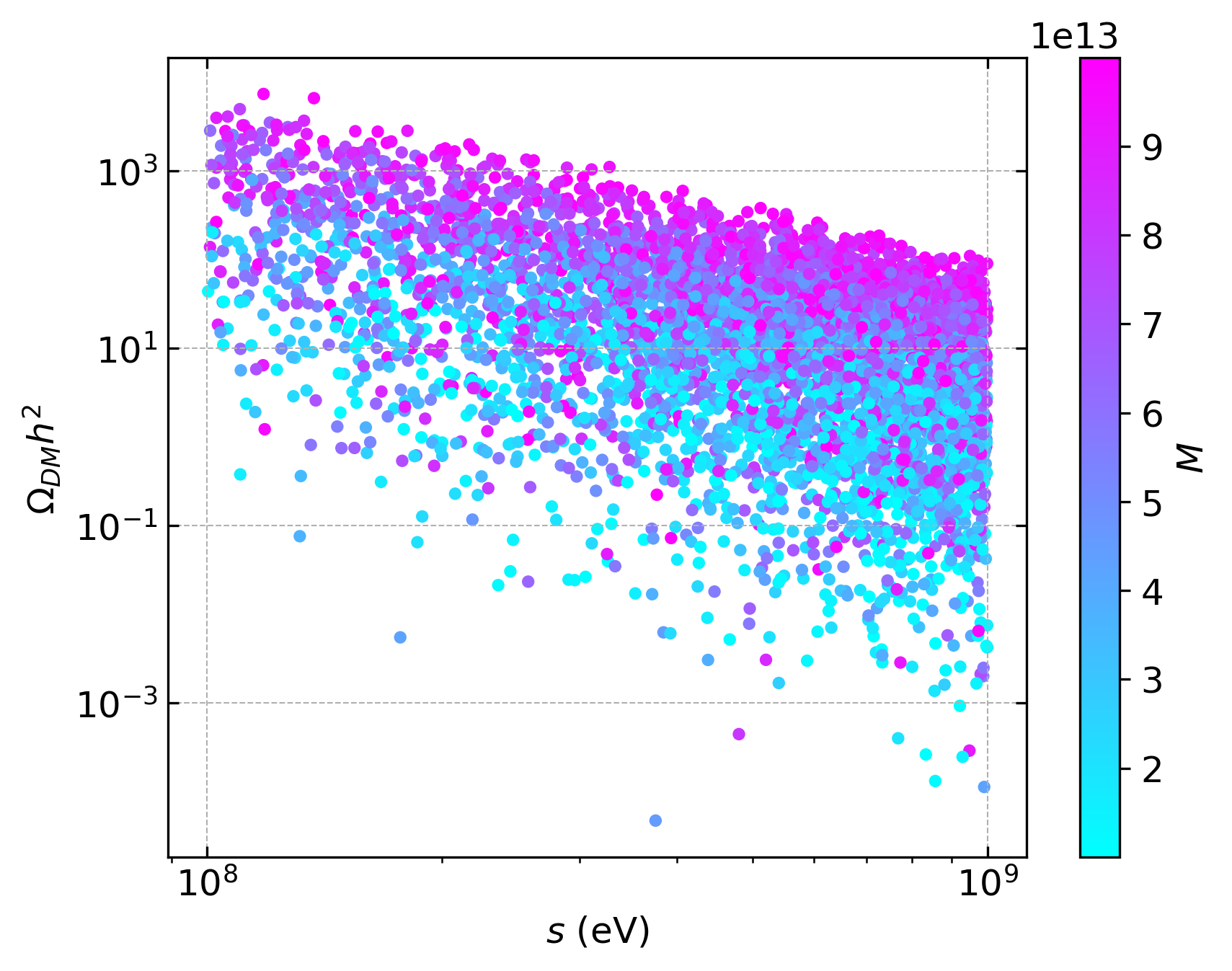}
     \end{subfigure}
     \hfill
     \begin{subfigure}{0.46\textwidth}
         \centering
\includegraphics[width=0.9\textwidth]{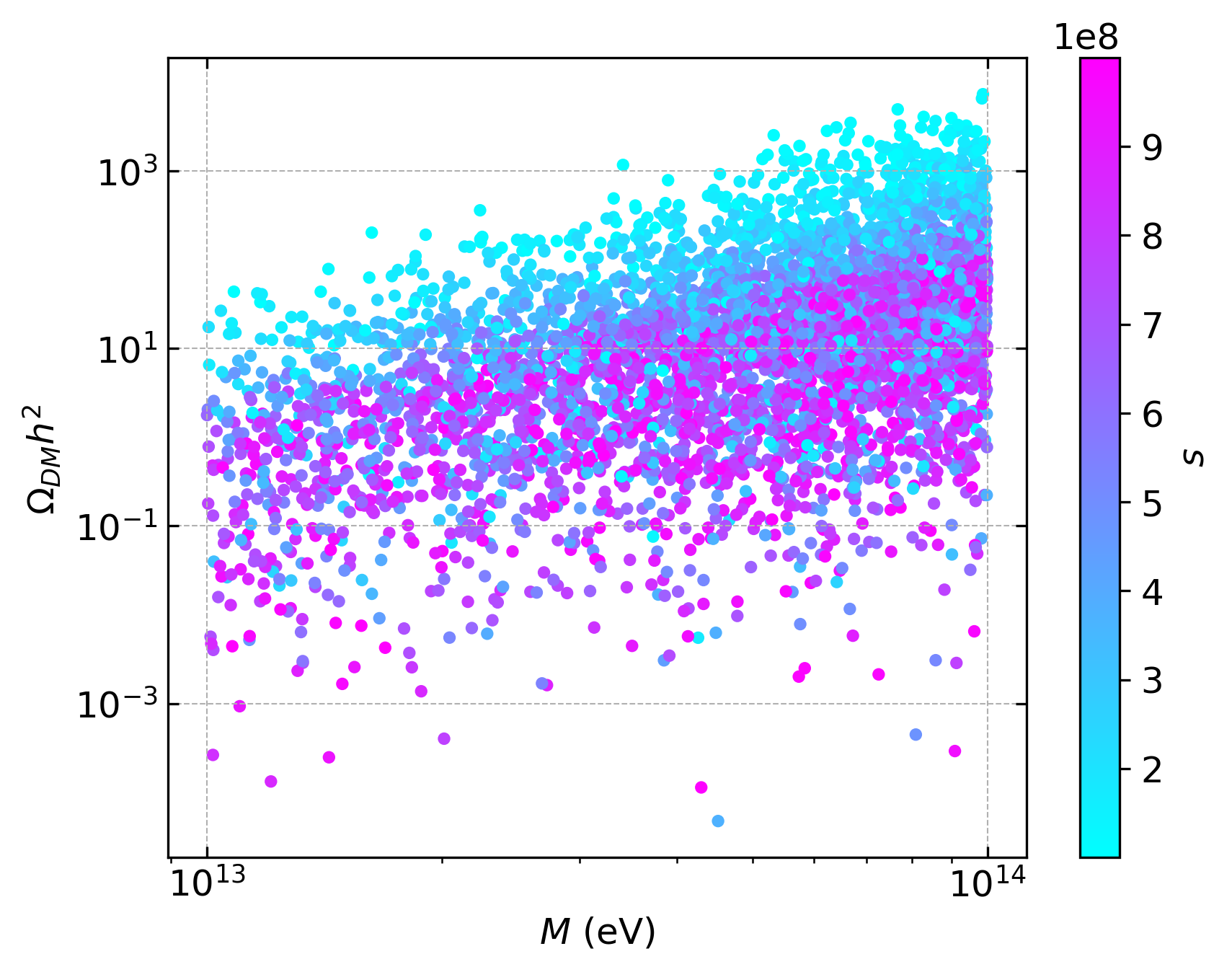}
     \end{subfigure}
     \end{figure}

     \begin{figure}
   \centering
\includegraphics[width=0.4\textwidth]{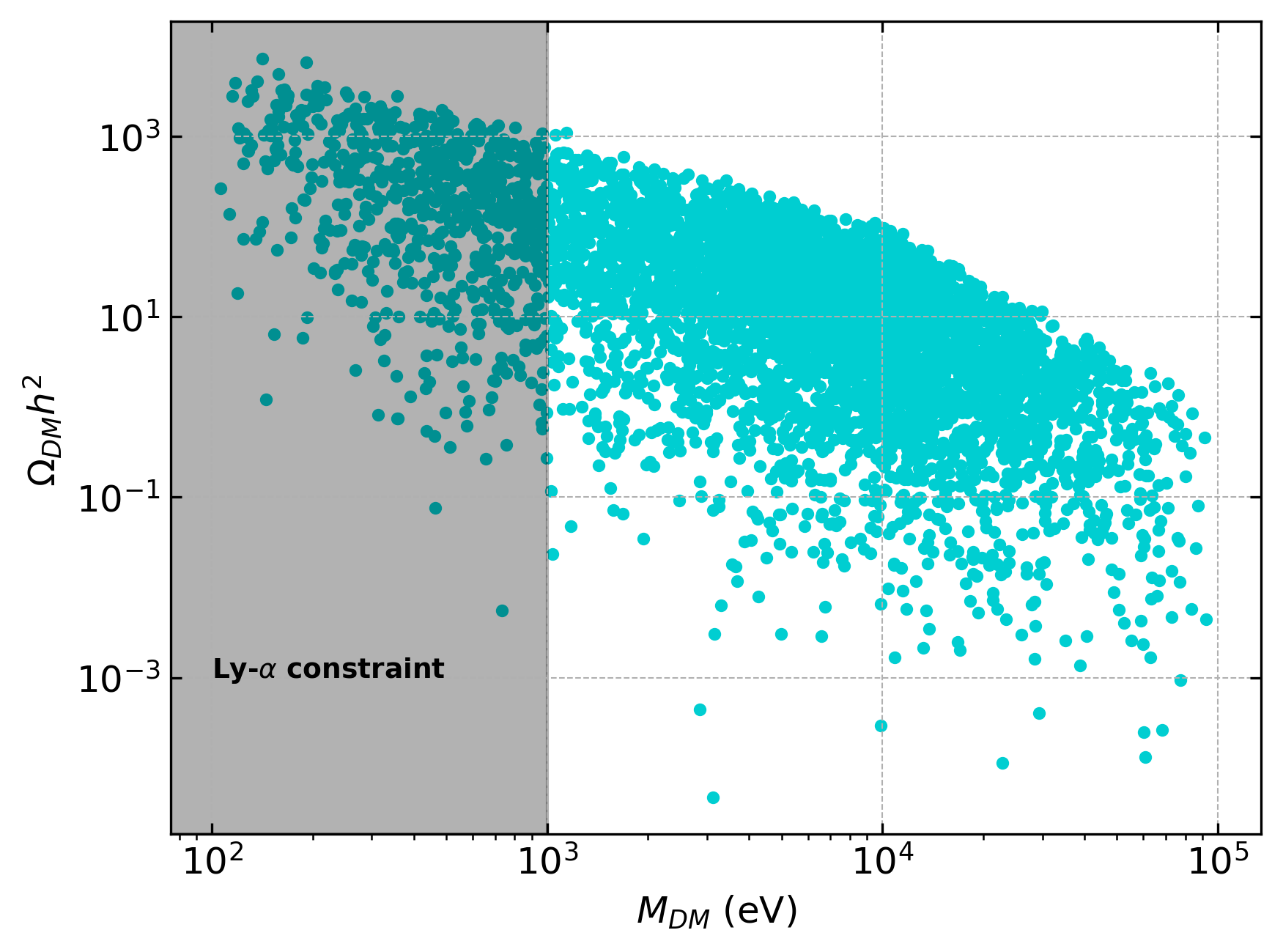}
\vspace{0.8cm}
    \caption{Plot in the first row shows  relic abundance ($\Omega_{DM} h^2$) as a function of Gauge singlet mass ($s$) and RHN mass ($M$) respectively, in the second row shows as a function of dark matter mass($M_{DM}$)  and  including constraints
from Lyman-$\alpha$ for NH.}
    \label{Fig7}   
\end{figure}

\section{Conclusion}
\label{conc}

We demonstrated the $\Delta(54)$ flavor symmetry along with the $Z_2 \otimes Z_3 \otimes Z_4$ symmetry which results in neutrino mass matrices and the subsequent predictions are consistent with the neutrino experimental data and the cosmological constraints on the relic abundance of dark matter (DM) and active neutrino-DM mixing angle. The ISS mechanism is incorporated into our model to offer a flavor-symmetric approach. This involves the prediction of the CP violation, solar mixing angle, upper octant of atmospheric mixing angle, and non-zero reactor angle.  The model best data
agreement is found in the normal hierarchy case with minimum $\chi^2$ analysis.   However, the inverted hierarchy scenario does not predict the model data with experimental findings. We calculated the sterile neutrino mass which is a potential dark matter (DM) candidate and the active neutrino-DM mixing angle with the predicted model parameters. The non-resonant formation of sterile neutrinos and the corresponding limitations from Lyman-$\alpha$ have been taken into consideration in this study. The active neutrino-DM mixing angle is examined in relation to DM mass. It is found that the data points from our model prefers the Lyman-$\alpha$ range in the case of NH. Future dark matter and neutrino oscillation experiments such as DUNE, JUNO, Daya Bay and Super-Kamiokande may test the model considering the predictions of the model align with the most recent neutrino data.

\section*{Acknowledgements} \par

HB acknowledges Tezpur University, India for Institutional Research Fellowship. The research of NKF is funded by DST-SERB, India under Grant no. EMR/2015/001683.

\bibliographystyle{naturemag} 
\bibliography{references}

\end{document}